\documentclass[aps,prd,twocolumn,groupedaddress,amssymb,eqsecnum,showpacs,epsfig]{revtex4}
\usepackage{graphicx}
\usepackage{bm}
\usepackage{dcolumn}
\usepackage{amsmath}
\def \lleq {\lower0.9ex\hbox{ $\buildrel < \over \sim$} ~}
\def \ggeq {\lower0.9ex\hbox{ $\buildrel > \over \sim$} ~}

\def \beq  {\begin{equation}}
\def \eeq  {\end{equation}}
\def \ber  {\begin{eqnarray}}
\def \eer  {\end{eqnarray}}

\def\apj{{Astroph.\@ J.\ }}

\def\aj{{Astron.\@ J.\ }}

\begin{document}

\newcommand{\newc}{\newcommand}

\newc{\be}{\begin{equation}}
\newc{\ee}{\end{equation}}
\newc{\ba}{\begin{eqnarray}}
\newc{\ea}{\end{eqnarray}}
\newc{\bea}{\begin{eqnarray*}}
\newc{\eea}{\end{eqnarray*}}
\newc{\D}{\partial}
\newc{\ie}{{\it i.e.} }
\newc{\eg}{{\it e.g.} }
\newc{\etc}{{\it etc.} }
\newc{\etal}{{\it et al.}}
\newcommand{\nn}{\nonumber}

\newc{\ra}{\rightarrow}
\newc{\lra}{\leftrightarrow}
\newc{\no}{Nielsen-Olesen }
\newc{\lsim}{\buildrel{<}\over{\sim}}
\newc{\gsim}{\buildrel{>}\over{\sim}}
\title{The fate of bound systems in phantom and quintessence cosmologies}
\author{S. Nesseris and L. Perivolaropoulos}
\email{http://leandros.physics.uoi.gr} \affiliation{Department of
Physics, University of Ioannina, Greece}
\date{\today}

\begin{abstract}
We study analytically and numerically the evolution of bound
systems in universes with accelerating expansion where the
acceleration either increases with time towards a Big Rip
singularity (phantom cosmologies) or decreases with time
(quintessence). We confirm the finding of Caldwell et. al.
\cite{Caldwell:2003vq} that bound structures get dissociated in
phantom cosmologies but we demonstrate that this happens earlier
than anticipated in Ref. \cite{Caldwell:2003vq}. In particular we
find that the `rip time' when a bound system gets unbounded is not
the time when the repulsive phantom energy gravitational potential
due to the average $(\rho + 3p)$ balances the attractive
gravitational potential of the mass M of the system. Instead, the
`rip time' is the time when the minimum of the time dependent
effective potential (including the centrifugal term) disappears.
For the Milky Way galaxy this happens approximately 180Myrs before
the Big Rip singularity instead of approximately 60Myrs indicated
in \cite{Caldwell:2003vq} for a phantom cosmology with w=-1.5. A
numerical reconstruction of the dissociating bound orbits is
presented.
\end{abstract}

\maketitle

\section{Introduction}
There is mounting observational evidence that the universe has
entered a phase of accelerating expansion (the scale factor obeys
${\ddot a}>0$) and that the total amount of clustered matter in
the universe is not sufficient for its small average spatial
curvature. This converging observational evidence comes from a
diverse set of cosmological data which includes observations of
type Ia supernovae \cite{snobs}, large scale redshift surveys
\cite{lss} and measurements of the cosmic microwave background
(CMB) temperature fluctuations spectrum \cite{cmb}. The observed
accelerating expansion and flatness of the universe, requires
either a modified theory of gravity\cite{modgrav} or, in the
context of standard general relativity, the existence of a smooth
energy component with negative pressure termed `dark
energy'\cite{dark energy}. This component is usually described by
an equation of state parameter $w\equiv{p\over \rho}$ (the ratio
of the homogeneous dark energy pressure $p$ over the energy
density $\rho$). For cosmic acceleration, a value of $w<-{1\over
3}$ is required as indicated by the Friedmann equation \be {{\ddot
a}\over a}=-{{4\pi G}\over 3}(\rho +3p) \label{fried}\ee Current
observational bounds \cite{snobs,phant-obs2} on the value of the
dark energy equation of state parameter $w(t_0)$ at the present
time $t_0$ yield \be -1.48<w(t_0)<-0.72 \ee at the $95 \%$
confidence level. The role of dark energy can be played by any
physical field with positive energy and negative pressure which
violates the strong energy condition $\rho +3p>0$ ($w>-{1\over
3}$).

Quintessence scalar fields\cite{quintess} ($-1<w<-{1\over 3}$)
violate the strong energy condition but not the dominant energy
condition $\rho + p
>0$. Their energy density scales down with the cosmic expansion
and so does the cosmic acceleration rate. Phantom
fields\cite{Caldwell:1999ew,Johri:2003rh,phantom} ($w<-1$) violate
the strong energy condition, the dominated energy condition and
maybe physically unstable. However, they are also consistent with
current cosmological data and according to recent
studies\cite{phant-obs2} they are favored over their quintessence
counterparts. In contrast to quintessence fields, the energy
density of phantom fields increases with time and so does the
predicted expansion acceleration rate ${{\ddot a}\over a}$. This
monotonically increasing acceleration rate of the expansion may be
shown to lead to a novel kind of singularity which occurs at a
finite future time and is characterized by divergences of the
scale factor $a$, the Hubble parameter $H$ its derivative ${\dot
H}$ and the scalar curvature. This singularity has been called
`Big Smash' \cite{McInnes:2001zw} the first time it was discussed
and `Big Rip' \cite{Caldwell:2003vq} (hereafter CKW) in a more
recent study. Even though there are mechanisms by which the `Big
Rip' singularity could be avoided \cite{Nojiri:2004ip} it remains
an interesting possible fate of the universe.

An immediate consequence of the very rapid expansion rate as the
Big Rip singularity is approached is the dissociation of bound
systems due to the buildup of repulsive negative pressure in the
interior of these systems. This observation was first made in
\cite{Caldwell:2003vq} where a qualitative study of the
dissociation times for various bound systems was also made.

The quantitative study however of the evolution of a bound system
in an expanding universe remains an issue of current research. In
particular the question of whether the expansion of the universe
affects gravitationally bound systems like clusters, galaxies or
planetary systems has been addressed in several early
\cite{ES,BS,Cal,Peeb} and recent
\cite{Baker:2001yc,Cooperstock:1998ny,Bon} studies. The recurrent
attention paid to this issue indicates that to this point a
definitive answer is still lacking. The prevalent perception
however is that the physics of systems which are small compared to
the radius of curvature of the cosmological background is
essentially unaffected by the expansion of the universe. Even
though this perception is applicable in the case of phantom
cosmology, it is not applicable as the Big Rip singularity is
approached and curvature increases rapidly.

A qualitative approach to the problem of bound system dissociation
in phantom cosmologies was made by CKW where it was assumed that a
bound system becomes unbound when the source of the repulsive
phantom energy gravitational potential for a bound system of size
$R$ $(-{4\over 3}\pi (\rho +3 p)R^3)$ balances the attractive
source of the ordinary gravitational potential (the mass $M$ of
the system). Even though this assumption is qualitatively useful
as it provides a physical understanding as to why does a bound
system dissociate due to phantom  energy repulsion, quantitatively
it leads to incorrect estimates of the dissociation times because
it ignores the effects of the centrifugal barrier in the effective
potential of bound systems. A quantitative analysis should make
use of the particle equations of motion in the local inertial
frame  based on the geodesic deviation equation. Using such an
equation, the time dependent effective potential of two body bound
systems may be shown to consist of three terms, the static
attractive gravitational mass term, the static repulsive
centrifugal term and the time-dependent repulsive dark energy
term. The stability of a bound system depends on the existence of
a minimum for the above effective potential. In the case of
quintessence $(w>-1)$ the minimum of the effective potential can
not disappear at any future time because the repulsive
time-dependent term decreases with time. For phantom energy
however $(w<-1)$ the repulsive time-dependent term increases with
time and at some critical time dominates over the other two terms
and destroys the minimum of the effective potential dissociating
at the same time the corresponding bound system. It should be
stressed that this critical time when the effective potential
minimum disappears is not the time when the repulsive dark energy
term balances the attractive gravitational mass term as assumed in
CKW. This will be demonstrated in detail, analytically and
numerically in the following sections.

The structure of this paper is the following: In section 2 we use
a metric that interpolates between the static Schwarzschild at
small scales and a general time dependent Friedmann universe
metric at large scales to derive the geodesics of a test particle
in the Newtonian limit. We then focus on the particular Friedmann
universe that contains a two component cosmic fluid (matter and
dark energy) and derive the form of the scale factor for any value
of constant $w$ (the equation of state parameter). Using this form
of the scale factor in the previously derived geodesic equation we
find the equation of motion for a two body bound system in an
expanding quintessence or phantom cosmological setup. In section 3
we study this equation of motion and derive the evolution of the
radius of two body bound systems with distance and mass scales
corresponding to the Solar System, the Milky Way Galaxy and the
Coma Cluster. We also derive analytically the dissociation time as
a function of $w$ for a bound system of given mass and radius and
test our result using numerical simulations of the above three
types of bound systems (Solar System, Milky Way and Coma Cluster).
Our analytical result for the time difference `Big Rip time -
Dissociation time' differs from the corresponding result of CKW by
a factor of 3.08 (we find the dissociation to occur earlier). The
time evolution of the bound system effective potential is also
shown. Our results are extended to the case of planar mass sources
and to the case of quintessence $(w>-1)$. In the later case it is
shown that the size change of the bound system due to the
expansion is negligible. Finally in section 4 we conclude,
summarize our main results and propose possible extensions of this
work.

\section{Geodesics in Expanding Schwarzschild Metric}

In order to investigate the effects of expansion on local bound
systems we must find the geodesics of test particles in the
appropriate metric that describes the spacetime in the vicinity of
a point mass $M$ placed in an expanding background. Such a metric
should interpolate between a static Schwarzschild metric at small
distances from $M$ and a time dependent Friedmann spacetime at
large distances. The detailed form of this interpolation is not
unique and there are different approaches to this problem in the
literature \cite{ES,BS,McV,Baker:2001yc}. In the Newtonian limit
(weak field, low velocities) such an interpolating metric takes
the form: \be ds^2=(1-\frac{2GM}{a(t)\rho})\cdot dt^2-a(t)^2\cdot
(d\rho^2+\rho^2\cdot (d\theta^2+sin^2\theta d\varphi^2))
\label{met} \ee

\noindent where $\rho$ is the comoving radial coordinate. Using
\be r=a(t) \cdot \rho \ee the geodesics corresponding to the line
element (\ref{met}) take the form \be -(\ddot{r}-{\ddot{a}\over
a}r)-{GM \over r^2}+r\dot{\varphi}^2=0 \label{geodr} \ee and \be
r^2\dot{\varphi}=L \label{geodf} \ee where $L$ is the constant
angular momentum per unit mass. Therefore the radial equation of
motion for a test particle in the Newtonian limit considered is
\be \ddot{r}={\ddot{a}\over a}r + {L^2 \over r^3}-{GM \over r^2}
\label{radeqm1} \ee

The same equation of motion is obtained in the Newtonian limit by
other interpolations even though the details in other limits may
vary \cite{Baker:2001yc}.

There is another simple and intuitive  (but not rigorous) way to
derive the same equation of motion by using Gauss's law for
gravity with gravitational sources the mass $M$ and the integral
of the homogeneous source $\rho+3p$. This approach leads to \be
{\ddot {\vec r}}=-{{GM}\over {r^2}}{\hat r}-{{4\pi G}\over
3}(\rho+3p){\vec r} \ee Using now the Friedmann equation
(\ref{fried}) this reduces to equation (\ref{radeqm1}).

Therefore the dynamics of a subluminal test particle bound in the
gravitational field of a mass $M$ (or equivalently a two body
bound system) in an expanding universe can be described by the
geodesic equation of motion (\ref{radeqm1}). In what follows we
will study the implications of this equation for two body bound
systems in various cosmologies.

As a warm up exercise let us consider the evolution of a bound
system in an expanding universe with scale factor \be a(t)\sim
t^\alpha \label{at1} \ee where $\alpha=const$. Let us assume that
at some initial time $t_0$ the test particle is at circular orbit
with radius $r_0$ and ${\dot \varphi}(t_0)=\omega_0={{GM}\over
{r_0^3}}$. Then the equation of motion (\ref{radeqm1}) may be
written in dimensionless form as \be {\ddot {\bar r}}-{{\bar
\omega_0}^2\over {{\bar r}^3}} + {{\bar {\omega_0}}^2\over {{\bar
r}^2}}-{{\alpha (\alpha -1)}\over {{\bar t}^2}}{\bar r}=0
\label{dleqm} \ee where ${\bar r}\equiv {r\over {r_0}}$, ${\bar
{\omega_0}}\equiv \omega_0 t_0$ and ${\bar t}\equiv {t\over t_0}$.
In what follows we will omit the bar (${\bar .}{\bar .}{\bar .}$)
for convenience but we shall work in dimensionless form. Typically
for gravitationally bound systems in the universe and cosmological
timescales we have \be \omega_0^2 = {{GM}\over {r_0^3}}t_0^2>>1
\ee (eg for galaxies $\omega_0 \simeq O(100)$) and we may
therefore consider the last term in equation (\ref{dleqm}) as a
perturbation perturbing the circular orbit $r_0=1$ to $r=1+\delta
r(t)$. It is then easy to show that (see also
\cite{Cooperstock:1998ny})  \be \delta r (t)={{\alpha (\alpha
-1)}\over {\omega_0^2 t^2}} \label{drpert} \ee  The radius of the
orbit tends to increase for $\alpha \in (0,1)$ (decelerating
expansion) while the perturbation $\delta r$ is negative because
the homogeneous {\it attractive} gravitational source ($\rho + 3p
> 0$ in this decelerating case) within a sphere of radius $r$
decreases with time. For $\alpha
> 1$ (accelerating expansion) the perturbation $\delta r$ is
positive but it is decreasing with time because the homogeneous
{\it repulsive} gravitational source ($\rho +3p < 0$ in this
accelerating case) within a sphere of radius $r$ decreases with
time.

To better understand physically the behavior of the perturbed
orbit we use the Friedmann equation (\ref{fried}). This equation
shows that the time dependent perturbing term of equation
(\ref{dleqm}) comes from the homogeneous gravitational source
$S={{4\pi}\over 3}(\rho +3p)r_0^3$ within the unperturbed radius
$r_0$. For an equation of state $p=w\rho$ we have \ba a&\sim &
t^{2\over {3 (w+1)}}=t^\alpha \label{at2}\\ \rho &\sim & a^{-3
(1+w)}\sim t^{-2} \label{rhot} \ea  The case $w > -{1\over 3}$
corresponds to $\alpha \in (0,1)$ and the radius perturbation
(\ref{drpert}) is negative and decreases with time. This is to be
expected because the gravitational source $S\sim \rho +3p$ is
positive (attractive) and decreases with time (equation
(\ref{rhot})). Thus $\delta r$ is negative (attractive
perturbation) but decreases with time (the energy density
decreases with time and so does the homogeneous gravitational
source).  For $-1<w<-{1\over 3}$ we have $\alpha > 1$ and the
radius perturbation (\ref{drpert}) is positive decreasing with
time. This is understood because the gravitational source $S$ is
negative (repulsive) and decreases with time. Thus $\delta r$ is
positive (repulsive perturbation) but decreases with time. Finally
for $w<-1$ we have $\alpha <0$. In this case the universe has an
expanding phase for $t<0$, a singularity (Big Rip) at $t=0$ and a
contracting phase at $t>0$. Our present expanding phase is
identified with the first phase ($t<0$) and the radius
perturbation is positive and {\it increasing} with time. This is
to be expected because the gravitational source $S$ is negative
(repulsive) but also increases with time as $t$ approaches the
singularity at $t=0$ (see equation \ref{rhot}). Thus $\delta r$ is
positive (repulsive perturbation) and increases with time.

This perturbative approach shows that the bound system radial
perturbation increases only in a phantom ($w<-1$, $\alpha < 0$)
expanding background but it can give no hint on whether a bound
system will eventually become unbound or not. A non-perturbative
approach is required to address this question in detail. Such an
approach should lead to the derivation of the full time evolution
of the radius of a two-body bound system. As discussed in the next
section this can be achieved by either explicitly solving the
equation of motion (\ref{radeqm1}) or by finding the time
evolution of the minimum of the effective potential derived from
equation (\ref{dleqm}).

Before proceeding with this more appropriate approach we will
improve on the estimate of the scale factor evolution by
considering a two-component (instead of one-component) fluid in a
Friedmann universe. Assuming the fluid components to be matter
with energy density $\rho_m$ and dark energy with density $\rho_x$
and equation of state $p_x=w\rho_x$ we may write the Friedmann
equations as \cite{Johri:2003rh}
\begin{equation}
\frac{{\dot a}^{2}}{a^2}\,=\, \frac{8\pi G}{3} \bigl[
\rho_m+\rho_x \bigr]
  \,=\,H_0^2 \bigl[ \Omega_m^0(a_0/a)^{3} +\Omega_x^0(a_0/a)^{3(1+w)}    \bigr]
\end{equation}
and \bea \frac{\ddot a}{a} &=&-\frac{4\pi
G}{3}[\rho_m+\rho_x(1+3w)]
=\,-\frac{4\pi G}{3}\, \rho_x \, [\Omega_x^{-1}+3w]  \nonumber\\
&=&-\frac{4\pi G}{3}\, \rho_x\,
           \left[ \frac{\Omega_m^0}{\Omega_x^0}
  \left( \frac{a_0}{a} \right)^{-3w}+1+3w \right]
\eea The transition from matter (decelerating) to dark energy
(accelerating) phase occurs at the transition redshift
\begin{equation}
1+z_m\,=\, {{a_0}\over {a_m}} \,=\, \Bigl[
\frac{-(3w+1)\Omega_x^0}{\Omega_m^0} \Bigr]^{-\frac{1}{3w}}
\end{equation}
At $z>z_m$ the solution of the Friedmann equation is
\begin{equation}
a^{3/2}(t) \,=\, \frac{3}{2}\xi t \label{at3}
\end{equation}
where $\xi$ is a constant. In the accelerating phase $z<z_m$ the
corresponding solution is approximated by
\begin{equation}
{a}^{3(1+w)/2}(t)\,=\,\frac{3}{2}\,\chi t + c \label{at4}
\end{equation}
where $c$ is an integration constant and $\chi = H_0
\sqrt{\Omega_x^0}$. Extending the expressions (\ref{at3}) and
(\ref{at4}) up to $t_m$ and matching for the evaluation of the
integration constants $\xi$ and $c$ leads to the expression for
the scale factor valid for $t>t_m$ ($z<z_m$)
\begin{equation}
a(t) \,=\,\frac{a(t_m)}{[-w+(1+w)t/t_m]^{-\frac{2}{3(1+w)}}}
\,\,\, for \;\; t >t_m \label{at5}
\end{equation}
For phantom energy ($w<-1$) the scale factor diverges at a finite
time \be t_*=\frac{w}{1+w}t_m >0 \label{brtime} \ee leading to the
Big Rip singularity. Since $\rho_x \sim a^{-3(1+w)}$ it is easy to
see that the phantom energy also diverges at $t_*$ as
\begin{equation}
\rho_x(t)\,=\,\frac{\rho(t_m)}{[-w+(1+w)t/t_m]^2}
\end{equation}
Using the expression (\ref{at5}) for the scale factor in the
two-component fluid universe we are in position to study in detail
the evolution of bound orbits in phantom and quintessence
cosmologies. This task will be undertaken in the following
section.

\section{Bound System Evolution}

Substituting the derived form of the scale factor in equation
(\ref{at5}) to the equation of motion (\ref{dleqm}) of a two-body
gravitating system in an expanding universe we obtain \be {\ddot
r}-{\omega_0^2\over
{r^2}}(1-\frac{1}{r})+\frac{2}{9}\frac{(1+3w)r}{(-w+(1+w)t)^2}=0
\label{dleqm1} \ee As in equation (\ref{dleqm}) this equation has
been made dimensionless by setting $\frac{r}{r_0}\rightarrow r$
and $\frac{t}{t_m}\rightarrow t$. The time dependent last term
proportional to \be \frac{\ddot a}{a}=-\frac{4\pi
G}{3}(\rho+3p)=\frac{2}{9} \frac{(1+3w)}{(-w+(1+w)t)^2} \ee
expresses the gravitating effects of the dark energy included
within a sphere of radius $r$ while the dimensionless parameter
$\omega_0^2$ is defined as \be \omega_0^2=\frac{GM}{r_0^3}t_m^2
\ee We shall use the dimensionless equation of motion
(\ref{dleqm1}) with initial conditions $r(t=1)=r_{min}$ and ${\dot
r}(t=1)\simeq 0$  ($r_{min}$ is the minimum of the effective
potential at $t=1$) to study the evolution of the radius of a two
body bound system which is initially at circular orbit. It will be
shown that in the phantom case ($w<-1$) the increasing repulsive
effects of the time dependent term of equation (\ref{dleqm1}) lead
to a dissociation of the bound system at a critical time
$t_{rip}$. A simplified qualitative approach to this question was
made by CKW where no reference was made to the equation of motion
as it was assumed that the dissociation occurs when the time
dependent dark energy gravitational source within the initial
radius balances the attractive gravitational source
$\frac{\omega_0^2}{r_{min}^2}$. This approach is only qualitative
and in many cases can lead to incorrect results for two reasons:
\begin{itemize}
\item It does not take into account the centrifugal force
$\frac{\omega_0^2}{r_{min}^3}$ \item It implicitly assumes that
the radius of the system just before dissociation is the same as
the initial radius $r_{min}$.
\end{itemize}
Nevertheless for comparison with our later quantitative exact
result we will rederive the result of CKW. The balance condition
of CKW may be written as \be \omega_0^2 \equiv
(\frac{2\pi}{T})^2=-\frac{2}{9}\frac{(1+3w)}{(-w+(1+w)t)^2} \ee
which leads to \be t_* - t_{rip} = \frac{T \sqrt{2\vert 1+3w
\vert}}{6\pi \vert 1+ w \vert} \label{cktrip} \ee where $t_*$ is
the Big Rip singularity time given by (\ref{brtime}). This is the
result of CKW to be compared with our quantitative result derived
in what follows.

The time-dependent effective potential that determines the
dynamics of the bound system is easily derived from equation
(\ref{dleqm1}) to be \be
V_{eff}=-\frac{\omega_0^2}{r}+\frac{\omega_0^2}{2r^2}-\frac{1}{2}
\lambda(t)^2 r^2 \label{veff}\ee  where \be
\lambda(t)=\frac{\sqrt{2\vert 1+3w\vert}}{3(-w+(1+w)t)}  \ee with
$w<-1$. At $t=1$ the system is assumed to be in circular orbit
with radius given by the minimum $r_{min}(t)$ of the effective
potential of equation (\ref{veff}). The location of $r_{min}(t)$
is time dependent and approximates the radius of the system at any
given time. It is the solution of the equation \be q(t)^2
r_{min}^4 =r_{min}-1 \label{rmineq} \ee where \be q(t)\equiv
\frac{\lambda(t)}{\omega_0} \ee It may be shown (using eg
Mathematica \cite{wolfram}) that this equation has a solution only
for \be q(t)^2 \leq \frac{27}{256}\equiv q_c \label{qcrit} \ee
Therefore the time $t_{rip}$ when the minimum of the potential
(\ref{veff}) disappears and the system becomes unbound is given by
the solution of the equation \be q(t_{rip})^2=\frac{27}{256}
\label{qcrit1} \ee It is straightforward to solve equation
(\ref{qcrit1}) for $t_{rip}$ and find \be t_* - t_{rip} = \frac{16
\sqrt{3}}{9} \frac{T \sqrt{2 \vert 1 + 3w \vert}}{6\pi \vert 1 + w
\vert} \label{trip1} \ee This result differs from the
corresponding result of CKW by the factor $\frac{16 \sqrt{3}}{9}
\simeq 3$.

\begin{figure}[h]
\centering
\includegraphics[bb=90 500 510 800,width=6.7cm,height=8cm,angle=0]{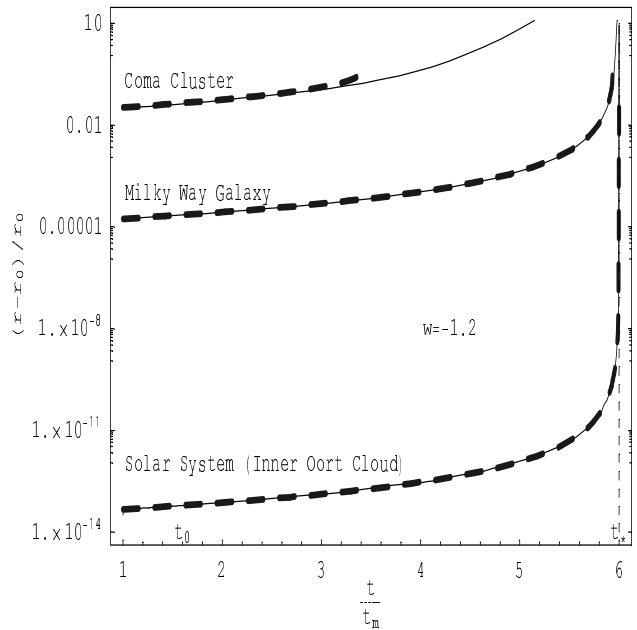}
\caption{The numerical evolution of the radius (continuous line)
and the analytical evolution of the effective potential minimum
(dashed line) for the three bound systems.} \label{fig1}
\end{figure}

In order to test this result numerically we must solve the
equation of motion (\ref{dleqm1}) for a specific bound system,
obtain numerically the orbits $r(t)$ and compare the analytical
prediction for the dissociation time with the corresponding time
visualized numerically. For concreteness we consider a phantom
cosmology with $w=-1.2$, $\Omega_m^0 = 0.3$ and $\Omega_x^0 =
0.7$. Our goal is to evaluate the dimensionless angular velocity
$\omega_0$ corresponding to specific systems in the context of
this cosmology and use it to calculate numerically the
corresponding future radial evolution. We will consider the scales
corresponding to three bound gravitational systems: the solar
system ($M=2\times 10^{33} gr$, $r_0=7\times 10^{15} cm$,
$\omega_0=3.5\times 10^6$), the Milky Way galaxy ($M=2\times
10^{45} gr$, $r_0=5\times 10^{22} cm$, $\omega_0=182$) and the
Coma Cluster ($M=6\times 10^{48} gr$, $r_0=9\times 10^{24} cm$,
$\omega_0=4.15$). In evaluating $\omega_0$ for the above systems
we have used the value of $t_m$ in the particular phantom
cosmology considered \be t_m=(H_0
\sqrt{\Omega_m^0})^{-1}\frac{2}{3} (1+z_m)^{-\frac{3}{2}} \simeq
1.8\times 10^{17} h^{-1} sec \ee We have evaluated the evolution
of the radius of the above systems using two methods: calculation
of the effective potential minimum (thick dashed lines of Fig.
\ref{fig1}) using equation (\ref{rmineq}) evolved until the
minimum disappears and explicit numerical evolution of the
equation of motion (\ref{dleqm1}) (continuous line) evolved up to
the dissociation time as obtained by CKW (equation
(\ref{cktrip})). The numerical evolution started at $t=t_m$ ($t=1$
in the dimensionless form) with initial orbit radius at the
minimum of the effective potential and negligible radial velocity
chosen such as to minimize radial oscillations. As seen in Fig.
\ref{fig1} there is
\begin{figure}[h]
\centering
\includegraphics[bb=85 530 475 795,width=7cm,height=8cm,angle=0]{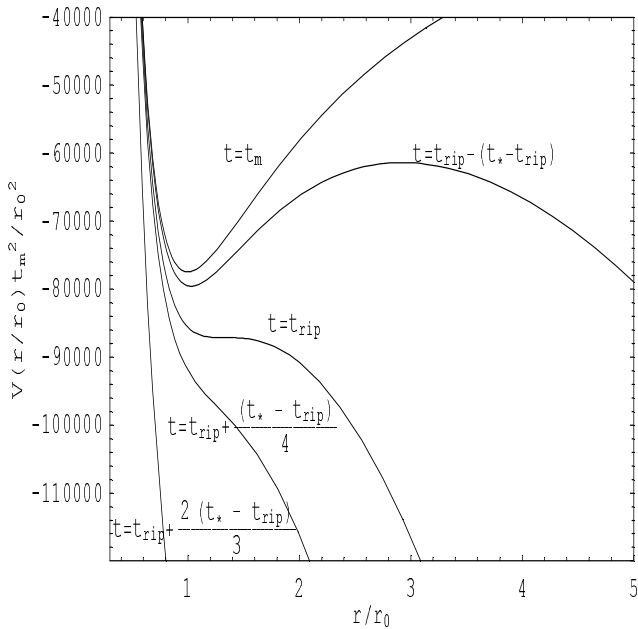}
\caption{The evolution of the dimensionless form of the effective
potential around the time trip for the Milky Way galaxy.}
\label{fig2}
\end{figure}
\noindent very good agreement between the numerical evolution of
the radius (continuous line) and the analytical evolution of the
effective potential minimum (dashed line) for the three bound
systems considered. The value of the present time $t_0$ is also
indicated on the time axis of Fig. \ref{fig1}. In Fig. \ref{fig2}
we show the evolution of the dimensionless form of the effective
potential around the time $t_{rip}$ when the minimum disappears
for a bound system corresponding to the Milky Way galaxy. As the
repulsive term destroys the minimum there is a small increase of
the location of the minimum and then a sudden disappearance and
dissociation of the system. The difference between our
quantitative prediction for $t_{rip}$ (indicated by the end of the
dashed lines in Fig. \ref{fig1}) and the corresponding qualitative
estimate of CKW (indicated by the end of the continuous lines) is
more prominent for the Coma Cluster and less so for the Milky Way.
The actual values in years for $t_* - t_{rip}$ are shown in Table
1 along with the qualitative prediction of CKW. The corresponding
dissociation times $t_{rip}$ in units of $t_m$ are shown in Table
2 \vspace{1cm}
\begin{center}
{\bf Table 1:} The dissociation times differences $t_*-t_{rip}$
for three bound systems in years as predicted by equations
(\ref{trip1}) and (\ref{cktrip}). The value $w=-1.2$ was assumed.
\begin{tabular}{|c|c|c|}\hline
{\bf System }& $t_*-t_{rip}$ (yrs)& $(t_*-t_{rip})_{CKW}$ (yrs) \\
\hline {\it Solar System }& $1.88\cdot 10^4 $& $6.11\cdot 10^3 $\\
\hline {\it Milky Way }& $3.59 \cdot 10^8$ & $1.17 \cdot 10^8$ \\
\hline {\it Coma Cluster }& $1.58 \cdot 10^{10}$ & $5.14 \cdot 10^9$ \\
\hline
\end{tabular}
\end{center}
For the case $w=-1.5$ considered by CKW, the corresponding
predictions for the Milky Way dissociation are $t_*-t_{rip}\simeq
166Myrs$ and $(t_*-t_{rip})_{CKW}\simeq 54Myrs$.

\begin{center}
{\bf Table 2:} The dissociation times $t_{rip}$ for the three
bound systems in units of $t_m$. For $w=-1.2$, $t_m\simeq 5.65
h^{-1} Gyrs$.

\begin{tabular}{|c|c|}\hline
{\bf System }& $t_{rip}/t_m$ \\
\hline {\it Solar System }& $6.00 $\\
\hline {\it Milky Way }& $5.94$  \\
\hline {\it Coma Cluster }& $3.19$ \\
\hline
\end{tabular}
\end{center}

Using the radial equation of motion (\ref{dleqm1}) along with the
conservation of angular momentum \be r^2 {\dot \varphi}=r_0^2
\omega_0 \label{angmom} \ee it is straightforward to obtain
numerically the full trajectory corresponding to the evolution of
the three bound

\begin{figure}[h]
\includegraphics[bb=70 330 520 790,width=8.0cm,height=8.0cm,angle=0]{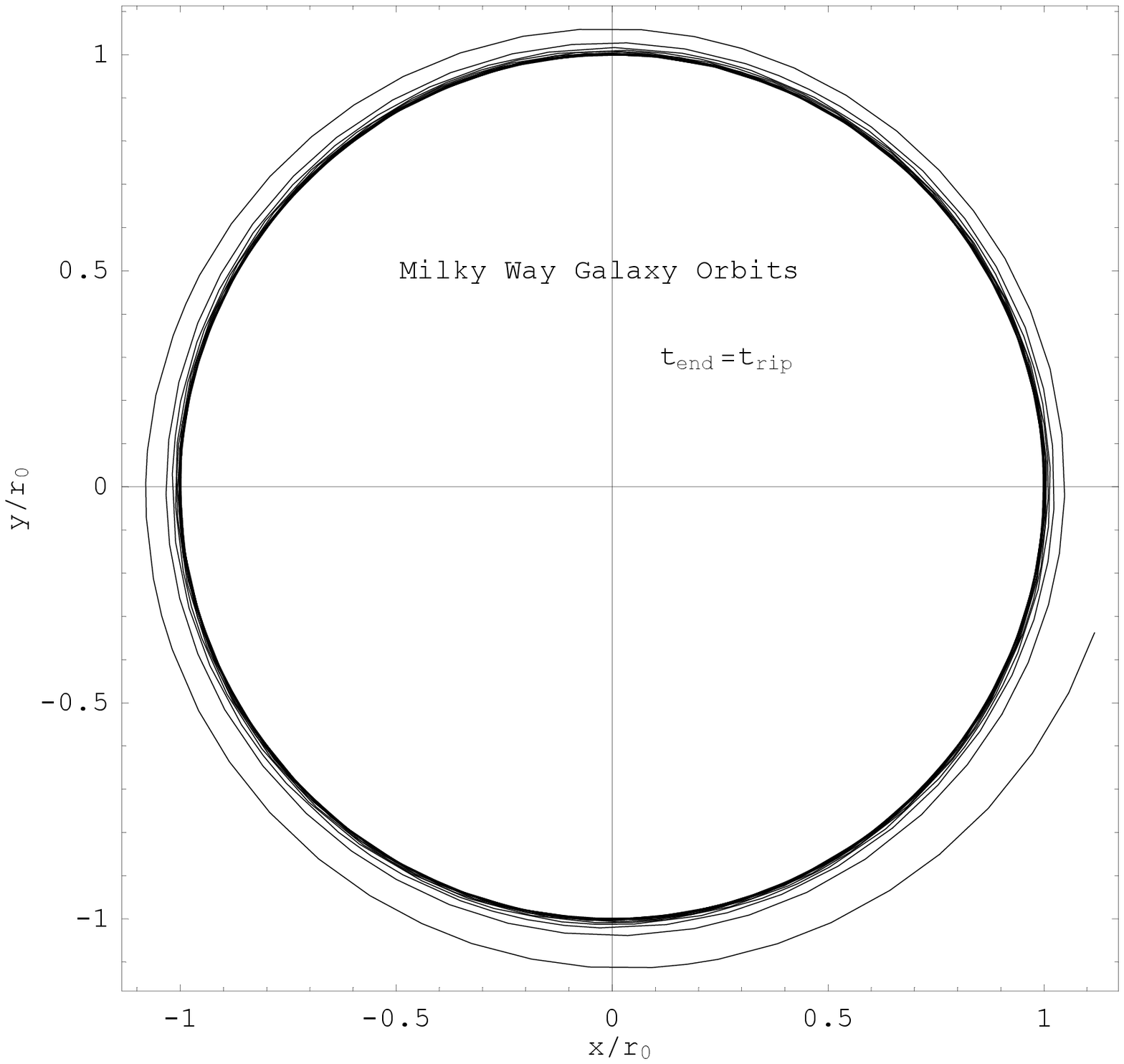}
\caption{The evolution of the the system of the Milky Way galaxy
until the effective potential minimum disappears.} \label{fig3}
\end{figure}
\begin{figure}[h]
\centering
\includegraphics[bb=85 350 500 800,width=7cm,height=7cm,angle=0]{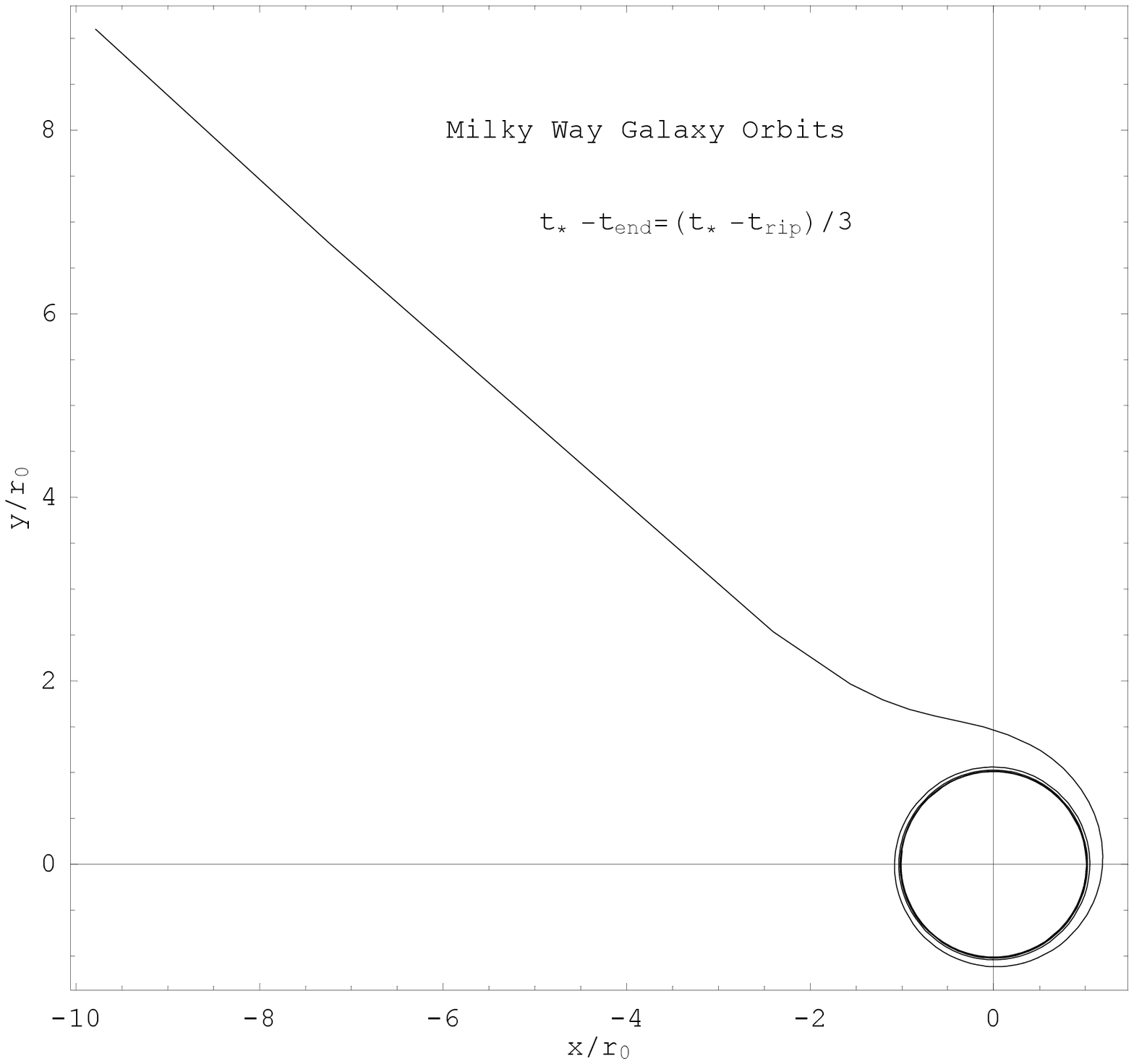}
\caption{The same system as in the previous figure but the
evolution lasts until the phantom energy balances the attractive
gravity of bound matter.} \label{fig4}
\end{figure}


\noindent systems and visualize the dissociation process. This
dissociation is demonstrated in Figs. \ref{fig3} and \ref{fig4}
for a two body bound system corresponding to the Milky Way galaxy.
In particular, in Fig. \ref{fig3} the system is evolved until the
effective potential minimum disappears while the evolution in Fig.
\ref{fig4} corresponds to the same system but lasts until the
phantom energy repulsive gravitational force balances the
attractive gravity of bound matter. Clearly the dissociation time
is the evolution time of Fig. \ref{fig3} while the evolution in
Fig. \ref{fig4} continuous well after the dissociation as
expected based on our analysis. Notice the radial (instead of
tangential) motion followed after dissociation which is due to the
dominant repulsive gravity of phantom energy.

\begin{figure}[h]\centering
\includegraphics[bb=75 495 540 805,width=7.5cm,height=7.0cm,angle=0]{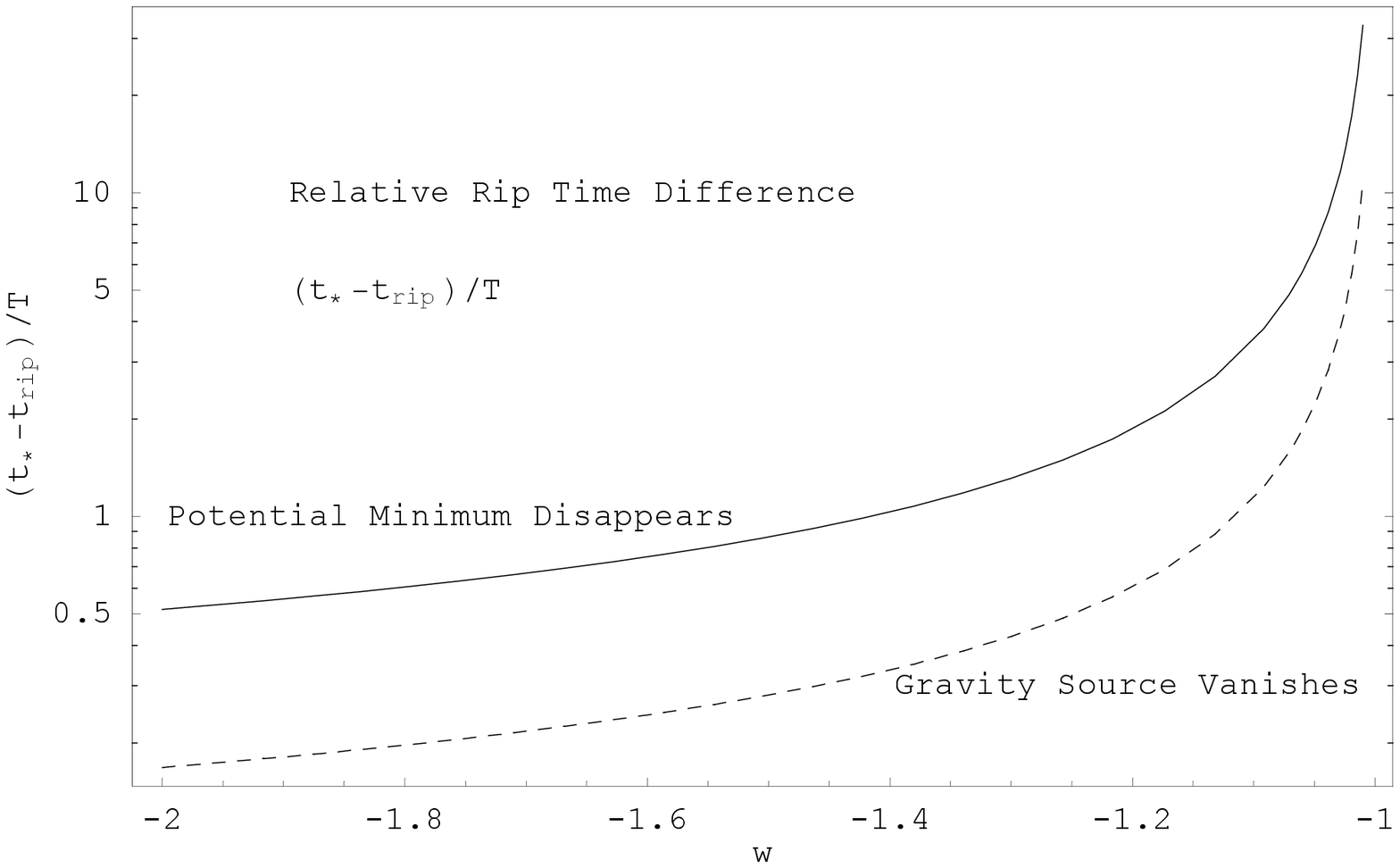}
\caption{The dependence on w of the relative rip time difference.}
\label{fig5}
\end{figure}

The numerical evolution shown in Figs. \ref{fig1} and \ref{fig2}
has assumed a phantom cosmology with $w=-1.2$. The dissociation
time $t_{rip}$ however is sensitive on the value of $w$ as shown
in equation (\ref{trip1}). To demonstrate this dependence Fig.
\ref{fig5} shows the dependence of the relative rip time
difference defined as \be \frac{t_* - t_{rip}}{T} \ee (where $T$
is the rotation period of the bound system) on $w$ for $w<-1$. The
continuous curve defines $t_{rip}$ as the time when the effective
potential minimum disappears while the corresponding definition
for the dashed curve is the time when the total gravity force
vanishes as in CKW. Clearly the two curves differ significantly
and the difference becomes more pronounced as $w$ approaches the
value $w=-1$.

\begin{figure}[h]
\centering
\includegraphics[bb=70 550 430 810,width=7cm,height=6.4cm,angle=0]{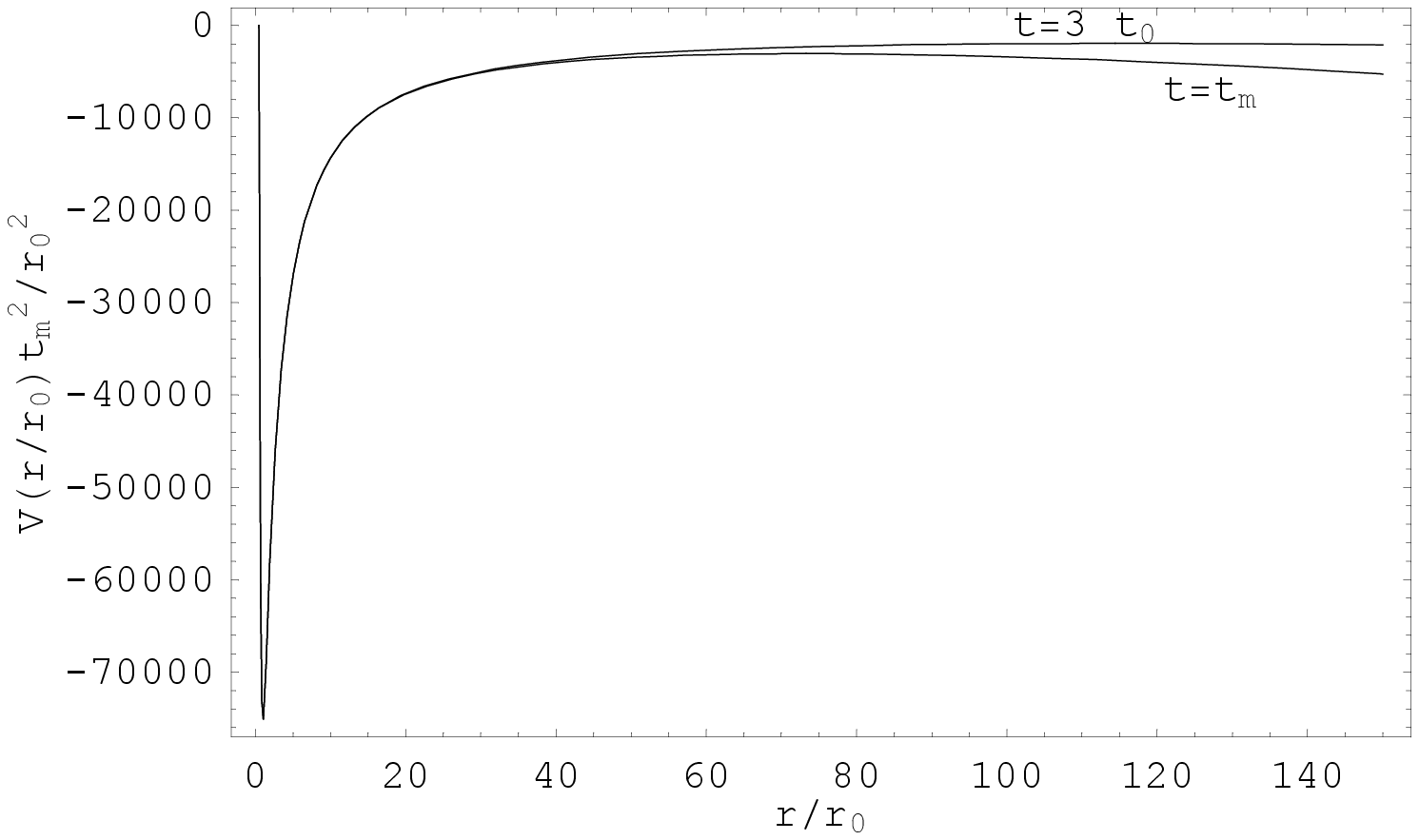}
\caption{The effective potential for the Milky Way galaxy with
$w=-0.9$ at times $t=t_m$ and $t=3 t_0$.} \label{fig6}
\end{figure}

\begin{figure}[h]
\includegraphics[bb=85 535 430 805,width=7cm,height=6.7cm,angle=0]{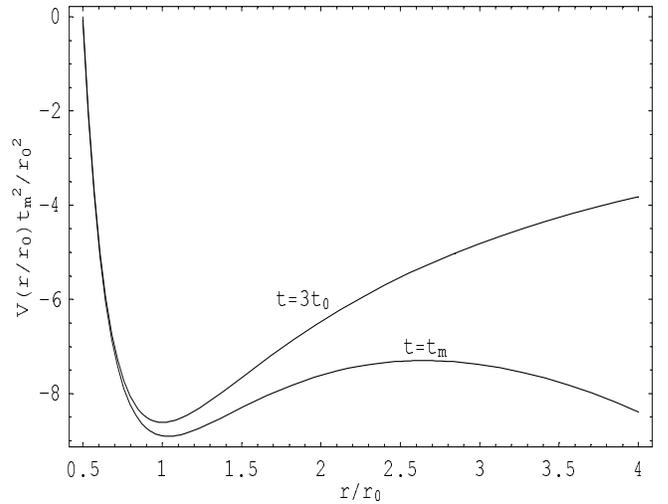}
\caption{The effective potential for the Coma Cluster with
$w=-0.9$ at times $t=t_m$ and $t=3 t_0$.} \label{fig7}
\end{figure}

Our discussion so far was based on the assumption of phantom
cosmologies ($w<-1$). This has been due to the fact that for
$w>-1$ (quintessence) the dark energy density decreases with time
and can therefore not destroy the effective potential minimum.
Quintessence can only cause negligible evolution (decrease) of the
radius of bound systems. This evolution can be easily obtained by
the perturbative treatment of section 2 by setting $\alpha =
\frac{2}{3(w+1)}$. To demonstrate the negligible effect of
expansion on bound systems for non-phantom cosmologies we have
plotted the effective potential for a Milky Way scale system with
$w=-0.9$ at times $t=t_m$ and $t=3 t_0$ (Fig. \ref{fig6}). The
corresponding plot for a Coma Cluster scale system is shown in
Fig. \ref{fig7}. The decrease of the radius is in both cases
minor but in the Coma Cluster case where $\omega_0$ is of $O(1)$
it is somewhat more prominent as expected from the perturbative
result (\ref{drpert}).

Finally it is of some interest to investigate the evolution of
bound systems with planar geometry. Consider a test particle at a
distance $h$ from a surface with surface density $\sigma$ in an
expanding universe background. Using Gauss's law for simplicity it
is straightforward to show that the equation of motion for the
particle is \be {\ddot h}=-2\pi G \sigma + \frac{3}{2} \frac{\ddot
a}{a} h \label{hdeq} \ee

\noindent Using equation (\ref{at5}) for the scale factor in a
phantom cosmology this may be written as \be t_m^2 {\ddot h} =
-h_0 +\beta h \label{eqh2} \ee where the derivative is with
respect to $\frac{t}{t_m}$, \be h_0=2\pi G \sigma t_m^2 \ee and
\be \beta(t) = \frac{\vert 1+3w \vert}{3(-w+\frac{t}{t_m}
(1+w))^2} \label{betadef} \ee Dividing by $h_0$ and setting
$\frac{t}{t_m} \rightarrow t$, $\frac{h}{h_0}\rightarrow h$,
equation (\ref{eqh2}) may be written in dimensionless form as \be
{\ddot h} =-1+\beta h \label{heq3} \ee The effective potential
corresponding to this equation has obviously no centrifugal term
and differs significantly

\begin{figure}[h]
\includegraphics[bb=130 515 485 755,width=7cm,height=6.4cm,angle=0]{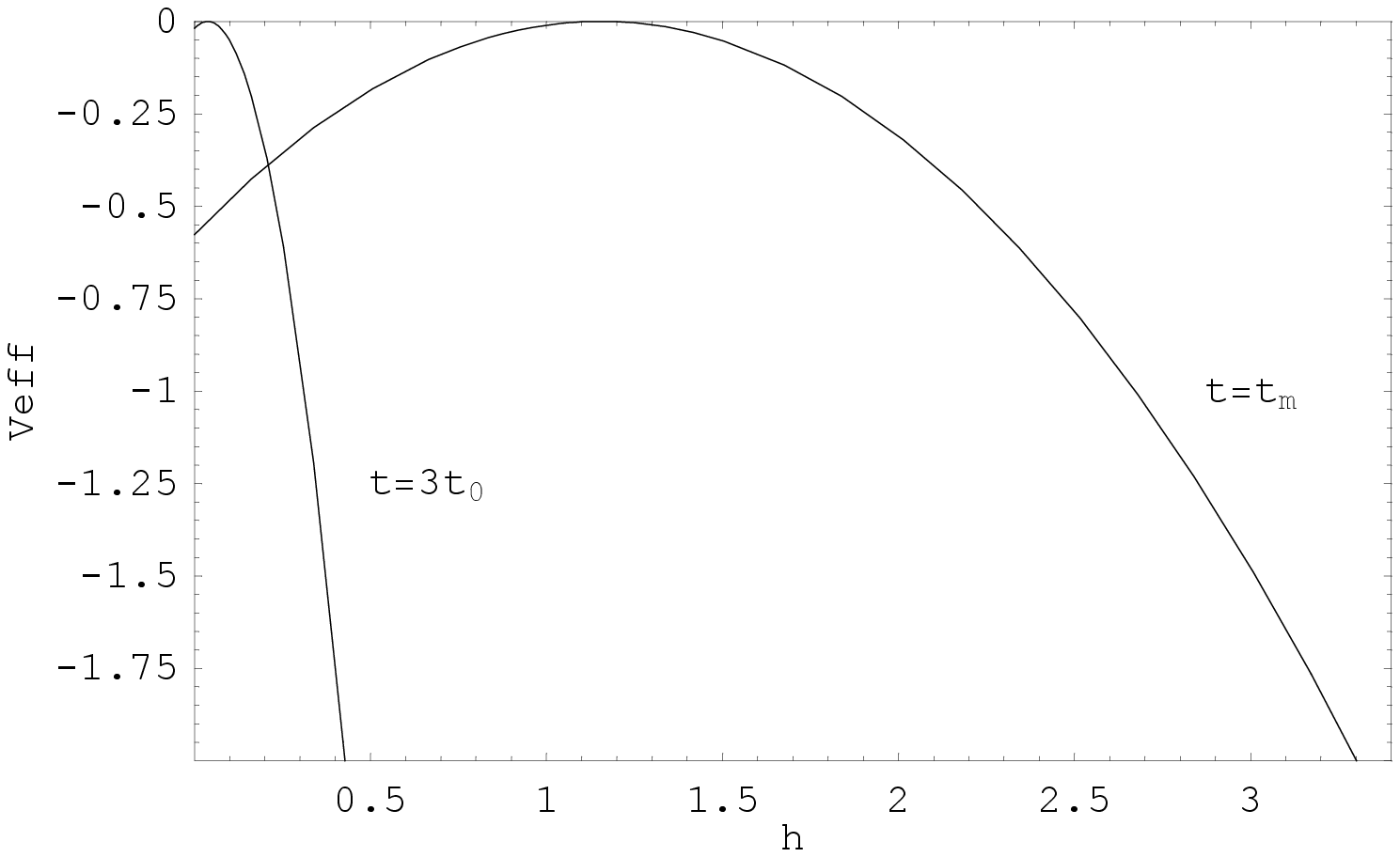}
\caption{The effective potential corresponding to bound systems
with planar geometry for $t=t_m,\; 3t_0$ and $w=-1.2$. There is
reflection symmetry of the potential with respect to $h=0$.}
\label{fig8}
\end{figure}

\noindent from the corresponding effective potential of the
spherically symmetric case. It is of the form \be V_{eff} =
-\frac{1}{2}\beta (h-\frac{1}{\beta})^2 \label{veff1} \ee and is
shown schematically in Fig. \ref{fig8}. It is a reversed harmonic
oscillator with a time-dependent unstable equilibrium point at
$h_{eq} = \frac{1}{\beta(t)}$. For quintessence $\beta(t)$
decreases with time and therefore the location of the equilibrium
point increases with time. Points initially on the left of the
equilibrium point will remain bound on the attractive side of the
potential. Points initially on the right of the equilibrium
(unbound) may also eventually end up bound on the left side of the
equilibrium point. For phantom cosmologies $\beta(t)$ increases
with time. This implies that the unstable equilibrium scale
$h_{eq}(t)$ decreases with time and therefore all scales will
eventually become larger than $h_{eq}(t)$ and dissociate. The
dimensionless scale $h_d (t_{rip})$ that dissociates at the time
$t_{rip}$ is found by solving the equation \be h_d =
\frac{1}{\beta(t_{rip})}=-\frac{3(-w+t_{rip} (1+w))^2}{1+3w}
\label{eqhd} \ee with solution \be t_*- t_{rip} =
\frac{T}{8\sqrt{3}} \frac{\sqrt{2\vert 1+ 3w \vert}}{\vert 1 + w
\vert } \label{htrip} \ee  where $T=4\sqrt{2h_d}$ is the period of
the oscillating test mass. Notice the similarity of this result
with the corresponding result found for spherically symmetric
systems (\ref{trip1}).

\section{Conclusion-Outlook}
We have studied the evolution of bound systems in expanding
backgrounds and focused on the case of accelerated expansion
powered by phantom energy ($w<-1$). We have found the radial time
dependence of bound systems in phantom cosmologies and determined
the time when these systems dissociate due to the repulsive
effects of phantom energy, as a function of the equation of state
parameter $w$. A universal behavior was found for the dissociation
time for different geometries of bound systems. We have also
plotted the bound system trajectory around the time of
dissociation and demonstrated that the bound systems explode
radially outward after dissociation. Our results were compared
with previous corresponding results in the literature and were
found to be in qualitative but not in quantitative agreement.

In the present study we have assumed a constant equation of state
parameter $w$. The extension of our results to the case of a
redshift dependent $w$ ($w(z)$) is straightforward and consists a
potentially interesting extension of this work. A potential
improvement to the accuracy of our results may come by using a
more accurate metric for the interpolation between the
Schwarzschild and the Friedmann metric \cite{Baker:2001yc}. Such
an improvement would be more important for strongly bound high
velocity systems.

The Mathematica\cite{wolfram} file used for the production of the
figures of the paper can be downloaded from \cite{mathfile} or
sent by e-mail upon request.

{\bf Acknowledgements:} This work was supported by the European
Research and Training Network HPRN-CT-2000-00152.


\begin{thebibliography}{99}

\bibitem{Caldwell:2003vq}
R.~R.~Caldwell, M.~Kamionkowski and N.~N.~Weinberg,
Phys.\ Rev.\ Lett.\  {\bf 91}, 071301 (2003)
[arXiv:astro-ph/0302506].

\bibitem{snobs}
Riess A \etal, 1998 \aj {\bf 116} 1009; Perlmutter S J \etal, 1999
\apj {\bf 517} 565;  S. Perlmutter et al, Nature (London)
{\bf391}, 51(1998); Bull.Am.Astron.Soc{\bf29},1351(1997);
P.Garnavich et al, Astrophys.J.{\bf493}L53 (1998); B.P.Schmidt et
al, Astrophys. J.{\bf507},46 (1998); Tonry, J L \etal, 2003 \apj
{\bf 594} 1; Barris, B \etal, 2004 \apj {\bf 602} 571; Knop R
\etal, 2003 \apj {\bf 598} 102; Riess A \etal, 2004 {\tt
astro-ph/0402512}.

\bibitem{lss}
N.A.Bahcall et al, Science{\bf284},1481 (1999); Percival, W.J.,
\etal, Mon.\@ Not.\@ Roy.\@ Ast.\@ Soc.\ {\bf 327}, 1297 (2001);
Max Tegmark et al, astro-ph/0310723.

\bibitem{cmb}
D.N.Spergel et al, arXiv.astro-ph/0302209; D.~Miller {\it et al.},
Astrophys.\ J.\  {\bf 524}, L1 (1999) [arXiv:astro-ph/9906421]; C.
Bennett et al, astro-ph/0302207;
P.~de Bernardis {\it et al.}  [Boomerang Collaboration],
Nature {\bf 404}, 955 (2000) [arXiv:astro-ph/0004404];
S.~Hanany {\it et al.},
Astrophys.\ J.\  {\bf 545}, L5 (2000) [arXiv:astro-ph/0005123];
T.~J.~Pearson {\it et al.},
Astrophys.\ J.\  {\bf 591}, 556 (2003) [arXiv:astro-ph/0205388];
B.~S.~Mason {\it et al.},
Astrophys.\ J.\  {\bf 591}, 540 (2003) [arXiv:astro-ph/0205384];
A.~Benoit {\it et al.}  [the Archeops Collaboration],
Astron.\ Astrophys.\  {\bf 399}, L25 (2003)
[arXiv:astro-ph/0210306].

\bibitem{modgrav} V. Sahni and Y. Shtanov, IJMP {\bf{D11}}, 1515 (2002); JCAP {\bf{0311}},
014 (2003); L.~Perivolaropoulos and C.~Sourdis,
Phys.\ Rev.\ D {\bf 66}, 084018 (2002) [arXiv:hep-ph/0204155]; C.
Deffayet, G.  Dvali and G. Gabadadze, Phys. Rev. {\bf{D65}},
044023 (2002); J. S. Alcaniz, Phys. Rev. D {\bf{65}}, 123514
(2002). astro-ph/0202492;  E. V. Linder, Phys. Rev. Lett.
{\bf{90}}, 091301 (2003); L.~Perivolaropoulos,
Phys.\ Rev.\ D {\bf 67}, 123516 (2003) [arXiv:hep-ph/0301237].

\bibitem{dark energy} M. S. Turner and M. White, Phys. Rev. {\bf{D56}}, R4439
(1997); A. Kamensshchik, U. Moschella and V. Pasquier, Phys. Lett.
B {\bf{511}}, 265 (2001); R. R. Caldwell, Phys. Lett. {\bf{B545}},
23 (2002); A. Dev, D. Jain and J. S. Alcaniz,  Phys. Rev.
{\bf{D67}}, 023515 (2003). astro-ph/0209379; J. S. Alcaniz, Phys.
Rev. {\bf{D69}}, 083521 (2004). astro-ph/0312424; 
M.~Axenides and K.~Dimopoulos,
JCAP {\bf 0407}, 010 (2004)[arXiv:hep-ph/0401238].; I.~Brevik, S.~Nojiri, S.~D.~Odintsov and L.~Vanzo,
Phys.\ Rev.\ D {\bf 70}, 043520 (2004)[arXiv:hep-th/0401073].; N.~J.~Nunes and D.~F.~Mota,
arXiv:astro-ph/0409481.; D.~F.~Mota and C.~van de Bruck, arXiv:astro-ph/0401504.

\bibitem{phant-obs2}
Alam U, Sahni V, Saini T D and Starobinsky A A, 2003 {\tt
astro-ph/0311364}; T. R. Choudhury, T. Padmanabhan,
astro-ph/0311622; O. Bertolami, A. A. Sen, S. Sen and P. T. Silva,
astro-ph/0402387; J. S. Alcaniz and N. Pires, Phys. Rev
{\bf{D70}}, 047303 (2004). astro-ph/0404146; Wang Y  and Mukherjee
P, 2004 \apj {\bf 606} 654; S.~Nesseris and L.~Perivolaropoulos,
Phys.\ Rev.\ D {\bf 70}, 043531 (2004) [arXiv:astro-ph/0401556];
Huterer D and Cooray A, 2004 {\tt astro-ph/0404062}; Gong Y, 2004
{\tt astro-ph/0401207}; Gong Y and Chen X, 2004 {\tt
gr-qc/0402031}; Gong Y, 2004 {\tt astro-ph/0405446}; Daly R A and
Djorgovski S G, 2004 {\tt astro-ph/0405550}; Corasaniti P S, Kunz
M, Parkinson D, Copeland E J and Bassett B A, 2004 {\tt
astro-ph/0406608}.

\bibitem{quintess}
B.Ratra and P.J.E.Peebles, Phys.Rev. D{\bf37},3406(1988);
astro-ph/0207347; C.Wetterich, Nucl. Phys. B{\bf302},668(1988);
P.G.Ferreira and M.Joyce, Phys. Rev. D.{\bf58},023503(1998);
P.Brax and J.Martin, Phys. Rev. D.{\bf61},103502(2000);
L.A.Urena-Lopez and T.Matos, Phys. Rev. D.{\bf62},081302(2000);
T.Barreiro, E.J.Copeland and N.J.Nunes, Phys. Rev.
D.{\bf61},127301(2000); D.Wands, E.J.Copeland and A.R.Liddle,
Phys. Rev. D.{\bf57},4686 (1998); A.R. Liddle and R.J. Scherrer,
Phys. Rev. D {\bf 59}, 023509 (1998); R.R. Caldwell, R. Dave and
P. Steinhardt, Phys. Rev. Lett {\bf 80}, 1582 (1998); I. Zlatev,
L. Wang and P. Steinhardt, Phys. Rev. Lett. {\bf 82}, 896 (1999);
S. Dodelson, M. Kaplinghat and E. Stewart, Phys. Rev. Lett. {\bf
85}, 5276(2000); V.B.Johri, Phys. Rev. D.{\bf63},103504(2001);
V.B.Johri, Class. Quant. Grav.{\bf19},5959 (2002); V.B.Johri,
Pramana {\bf59},1(2002); T.D.Saini, S. Raychaudhury, V. Sahni and
A. A. Starobinsky, Phys. Rev. Lett{\bf85},1162(2000); V.Sahni and
L.Wang, Phys. Rev. D.{\bf62},103517(2000).

\bibitem{Caldwell:1999ew}
R.~R.~Caldwell,
Phys.\ Lett.\ B {\bf 545}, 23 (2002) [arXiv:astro-ph/9908168];
P.~F.~Gonzalez-Diaz,
Phys.\ Lett.\ B {\bf 586}, 1 (2004) [arXiv:astro-ph/0312579];
P.~Singh, M.~Sami and N.~Dadhich,
Phys.\ Rev.\ D {\bf 68}, 023522 (2003) [arXiv:hep-th/0305110].

\bibitem{Johri:2003rh}
V.~B.~Johri,
Phys.\ Rev.\ D {\bf 70}, 041303 (2004) [arXiv:astro-ph/0311293].

\bibitem{phantom}
A.E.Schulz and Martin White, Phys. Rev. D{\bf64},043514(2001);
V.K.Onemli and R.P.Woodard, Class. Quant. Grav.{\bf19}, 4607
(2000); V.K.Onemli and R.P.Woodard, gr-qc/0406098; D.F.Torres,
Phys. Rev. D{\bf66},043522(2002); P.Singh, M.Sami and N.Dadhich,
arXiv.hep-th/0305110; S. Hannestad and E. Mortsell, Phys. Rev.
D{\bf66},063508(2002); A.Melchiorri, L.Mersini, C.J.Odman and
M.Trodden, astro-ph/0211522; S.M.Carroll, M.Hoffman and M.Trodden,
astro-ph/0301273; P.H.Frampton, hep-th/0302007; Jian-gang Hao and
Xin-zhou Li, hep-th/0302100, hep-th/0303093, hep-th/0305207;
G.W.Gibbons, hep-th/0302199; Shin'ichi Nojiri and S.D.Odintsov,
hep-th/0303117, hep-th/0304131; J.A.S.Lima, J.V.Cunha and S.
Alcanz, astro-ph/0303388; A.Yurov, astro-ph/0305019; B.Mcinnes,
hep-th/0305107, astro-ph/0210321; P.F.Gonzalez-Diaz,
astro-ph/0305559; M.P.Dabrowski, T.Stachowiak and Marck
Szydlowski, astro-ph/0307128; 
B.~Boisseau, G.~Esposito-Farese, D.~Polarski and A.~A.~Starobinsky,
Phys.\ Rev.\ Lett.\  {\bf 85}, 2236 (2000)[arXiv:gr-qc/0001066]; L.~P.~Chimento and R.~Lazkoz,
Phys.\ Rev.\ Lett.\  {\bf 91}, 211301 (2003)[arXiv:gr-qc/0307111];
H.~Stefancic,Phys.\ Lett.\ B {\bf 586}, 5 (2004)
[arXiv:astro-ph/0310904];H.~Stefancic,Eur.\ Phys.\ J.\ C {\bf 36}, 523 (2004)
[arXiv:astro-ph/0312484];P.~F.~Gonzalez-Diaz,
Phys.\ Rev.\ D {\bf 69}, 063522 (2004)[arXiv:hep-th/0401082];P.~F.~Gonzalez-Diaz,
Phys.\ Rev.\ Lett.\  {\bf 93}, 071301 (2004)[arXiv:astro-ph/0404045];

\bibitem{McInnes:2001zw}
B.~McInnes,
JHEP {\bf 0208}, 029 (2002) [arXiv:hep-th/0112066].

\bibitem{Nojiri:2004ip}
S.~Nojiri and S.~D.~Odintsov,
Phys.\ Lett.\ B {\bf 595}, 1 (2004)
[arXiv:hep-th/0405078]; E.~Elizalde, S.~Nojiri and S.~D.~Odintsov,
Phys.\ Rev.\ D {\bf 70}, 043539 (2004)
[arXiv:hep-th/0405034]; S.~Nojiri and S.~D.~Odintsov,
arXiv:hep-th/0408170; P.~X.~Wu and H.~W.~Yu,
arXiv:astro-ph/0407424.

\bibitem{ES} A. Einstein and E. G. Straus, Rev.\ Mod.\ Phys.\ {\bf 17},
120 (1945); erratum {\bf 18}, 148 (1946).

\bibitem{BS} C. Bona and J. Stela, Phys.\ Rev.\ D {\bf 36}, 2915 (1987).

\bibitem{Cal} C. Callan  et al., Am.\ J.\ Phys.\ {\bf 33},
105 (1965).

\bibitem{Peeb} R. H. Dicke and P. J. E. Peebles, Phys.\ Rev.\ Lett\ {\bf 12},
435 (1964).

\bibitem{Baker:2001yc}
G.~A.~Baker,
[arXiv:astro-ph/0112320].

\bibitem{Cooperstock:1998ny}
F.~I.~Cooperstock, V.~Faraoni and D.~N.~Vollick,
Astrophys.\ J.\  {\bf 503}, 61 (1998) [arXiv:astro-ph/9803097].

\bibitem{Bon}
W. B. Bonnor, M. N. R. A. S. {\bf 282}, 1467 (1996);
H.~Stefancic,
Phys.\ Lett.\ B {\bf 595}, 9 (2004)[arXiv:astro-ph/0311247];P.~F.~Gonzalez-Diaz and C.~L.~Siguenza,
Phys.\ Lett.\ B {\bf 589}, 78 (2004).

\bibitem{McV} G. C. McVittie, M. N. R. A. S. {\bf 93}, 325 (1933).

\bibitem{wolfram} http://www.wolfram.com/

\bibitem{mathfile} http://leandros.physics.uoi.gr/bigrip.htm


\end{thebibliography}
\end{document}